\documentclass[prb,fleqn,12pt,showpacs]{revtex4}
\usepackage{epsfig}
\usepackage{graphicx}
\usepackage{amsfonts}
\begin{document}
\title{Orbital fluctuations, spin-orbital coupling, and anomalous \\
magnon softening in an orbitally degenerate ferromagnet}
\author{Dheeraj Kumar Singh, Bhaskar Kamble, and Avinash Singh}
\email{avinas@iitk.ac.in}
\affiliation{Department of Physics, Indian Institute of Technology Kanpur}
\begin{abstract}
The correlated motion of electrons in the presence of strong orbital fluctuations and correlations is investigated with respect to magnetic couplings and excitations in an orbitally degenerate ferromagnet within the framework of a non-perturbative Goldstone-mode-preserving approach based on a systematic inverse-degeneracy expansion scheme. Introduction of the orbital degree of freedom results in a class of diagrams representing spin-orbital coupling which become particularly important near the orbital ordering instability. Low-energy staggered orbital fluctuation modes, particularly with momentum near $(\pi/2,\pi/2,0)$ (corresponding to period $4a$ orbital correlations as in CE phase of manganites involving staggered arrangement of nominally $\rm Mn^{3+} / \rm Mn^{4+}$ ions, and staggered ordering of occupied $3x^2 - r^2 / 3y^2 - r^2$ orbitals on alternating $\rm Mn^{3+}$ sites), are shown to generically yield strong intrinsically non-Heisenberg $(1-\cos q)^2$ magnon self energy correction, resulting in no spin stiffness reduction, but strongly suppressed zone-boundary magnon energies in the $\Gamma$-X direction. The zone-boundary magnon softening is found to be strongly enhanced with increasing hole doping and for narrow-band materials, which provides insight into the origin of zone-boundary anomalies observed in ferromagnetic manganites.

\end{abstract}
\maketitle
\newpage
\section{Introduction}
The orbital degree of freedom of the electron has attracted considerable attention in recent years due to the rich variety of electronic, magnetic, and transport properties exhibited by orbitally degenerate systems such as the ferromagnetic manganites, which have highlighted the interplay between spin and orbital degrees of freedom in these correlated electron systems.\cite{tokura_2003,khaliullin_2005} Orbital fluctuations, correlations, and orderings have been observed in Raman spectroscopic studies\cite{saitoh_2001} of orbiton modes in $\rm LaMnO_3$, polarization-contrast-microscopy studies\cite{ogasawara_2001} of $\rm La_{0.5}Sr_{1.5}MnO_4$, magnetic susceptibility and inelastic neutron scattering studies\cite{khalifah_2002} of $\rm La_4 Ru_2 O_{10}$, and resonant inelastic soft X-ray scattering studies\cite{ulrich_2008} of $\rm YTiO_3$ and $\rm LaTiO_3$. A new detection method for orbital structures and ordering based on spectroscopic imaging scanning tunneling microscopy is of strong current interest\cite{lee_2009} in orbitally active metallic systems such as strontium ruthenates and iron pnictide superconductors.

The role of orbital fluctuations on magnetic couplings and excitations is of strong current interest in view of the several zone-boundary anomalies observed in spin-wave excitation measurements in the metallic ferromagnetic phase of colossal magnetoresistive (CMR) manganites.\cite{hwang_98,dai_2000,tapan_2002,ye_2006,ye_2007,zhang_2007,moussa_2007} The presence of short-range dynamical orbital fluctuations has been suggested in neutron scattering studies of ferromagnetic metallic manganite $\rm La_{1-x}(Ca_{1-y}Sr_y)_{x}MnO_3$.\cite{moussa_2007}
These observations are of the crucial importance for a quantitative understanding of the carrier-induced spin-spin interactions, magnon excitations, and magnon damping, and have highlighted possible limitations of existing theoretical approaches. 

For example, the observed magnon dispersion in the $\Gamma$-X direction shows significant softening near the zone boundary, indicating non-Heisenberg behaviour usually modeled by including a fourth neighbour interaction term $J_4$, and highlighting the limitation of the double-exchange model. Similarly, the prediction of magnon-phonon coupling as the origin of magnon damping\cite{dai_2000} and of disorder as the origin of zone-boundary anomalous softening\cite{furukawa_2005} have been questioned in recent experiments.\cite{ye_2006,ye_2007,zhang_2007,moussa_2007} Furthermore, the dramatic difference in the sensitivity of long-wavelength and zone-boundary magnon modes on the density of mobile charge carriers has emerged as one of the most puzzling feature. Observed for a finite range of carrier concentrations, while the spin stiffness remains almost constant, the anomalous softening and broadening of the zone-boundary modes show substantial enhancement with increasing hole concentration.\cite{ye_2006,ye_2007}

Theoretically, the role of orbital-lattice fluctuations and correlations on magnetic couplings and excitations has been  investigated within an orbitally degenerate double exchange model with an inter-orbital interaction $V$ and the Jahn-Teller coupling.\cite{khaliullin_2000} Based on a strong coupling expansion, this approach is restricted to the strong coupling limit $V \gg t$. The final calculations for the magnon self energy, carried out in terms of a phenomenological parameter, show significant zone-boundary magnon softening only for ferromagnetic orbital correlations, and extremely close to the orbital ordering instability. 


If orbital fluctuations have signature effects on magnetic excitations in a ferromagnet with orbital degree of freedom, they can be probed indirectly through neutron scattering studies. A detailed investigation of the orbital fluctuation magnon self energy is therefore of strong current interest, especially with respect to dependence on inter-orbital interaction strength, band filling, and different orbital fluctuation modes. In this paper we will present a theory for spin-orbital coupling and magnon self energy, and examine how the correlated motion of electrons in the presence of strong orbital correlations near the orbital ordering instability influences magnetic couplings and excitations. 

We will employ a diagrammatic approach which allows interpolation in the full range of interaction strength from weak to strong coupling. In this approach correlation effects in the form of self energy and vertex corrections are incorporated systematically so that the Goldstone mode is preserved order by order. Based on a systematic inverse-degeneracy expansion scheme,\cite{singh} the approach has been used recently to study spin-charge coupling effects, which give rise to strong magnon energy softening, damping, and non-Heisenberg behaviour in metallic ferromagnets.\cite{spch3,qfklm} 

The present work will also extend the recent investigation into role of orbital degeneracy and Hund's coupling on magnetic couplings and excitations in a band ferromagnet.\cite{hunds} Orbital degeneracy and Hund's coupling were shown to enhance ferromagnetism by strongly suppressing correlation-induced quantum corrections to spin stiffness and magnon energies. An effective quantum parameter was obtained for determining the magnitude of quantum corrections, and the theory was applied to calculate the spin stiffness for a realistic multi-orbital system such as iron. We will show here that the spin stiffness remains essentially unaffected by the interaction $V$ due to the non-Heisenberg $(1-\cos q)^2$ behaviour of the magnon self energy resulting from orbital fluctuations and correlations.  

In manganites, an important role is also played by the cooperative Jahn-Teller distortion of O$^{2-}$ ions which lifts the two-fold degeneracy of $\rm e_g$ electronic levels of Mn due to a combination of orbital geometry and electrostatic repulsion, leading to staggered orbital correlations. This is qualitatively similar to the local orbital moment and staggered orbital correlations introduced by the inter-orbital density interaction $V n_{i\alpha} n_{i\beta}$ which relatively pushes up the $\beta$ orbital energy if the $\alpha$ orbital density $\langle n_{i\alpha} \rangle$ is more than average, thus self consistently lifting the orbital degeneracy. Therefore, orbital correlations and fluctuations due to dynamical Jahn-Teller distortion can be qualitatively treated in terms of an effective inter-orbital interaction. 

The structure of the paper is as follows. Starting with a degenerate two-orbital Hubbard model including an inter-orbital interaction $V$, the first order quantum correction diagrams for the irreducible particle-hole propagator are obtained in Section II. As basic ingredients in the diagrammatics, spin and orbital fluctuations are briefly discussed in section III. The magnon self energy contributions due to orbital fluctuations and spin-orbital coupling are obtained in sections IV and V. The interplay between magnetic and charge contributions to the spin-orbital interaction vertex is discussed in section VI, and orbital fluctuations near $(\pi/2,\pi/2,0)$ are shown to yield strong zone-boundary magnon softening. Extension to finite Hund's coupling $J$ and the ferromagnetic Kondo lattice model are discussed in sections VII and VIII, and conclusions are presented in Section IX. 

\section{Two-orbital Hubbard model}
We will consider a two-orbital Hubbard model 
\begin{eqnarray}
H &=& -t \sum_{\langle ij\rangle \sigma} (a^\dagger _{i\alpha\sigma} a_{j\alpha\sigma} + a^\dagger _{i\beta\sigma}  a_{j\beta\sigma} ) + U \sum_i (n_{i\alpha\uparrow} n_{i\alpha\downarrow} + n_{i\beta\uparrow} n_{i\beta\downarrow} ) \nonumber \\ &+& \sum_{i\sigma\sigma'} (V-\delta_{\sigma\sigma'}J) n_{i\alpha\sigma} n_{i\beta\sigma'} 
- J \sum_{i,\sigma \ne \sigma'} a^\dagger _{i\alpha\sigma} a_{i\alpha\sigma'} a^\dagger _{i\beta\sigma'} a_{i\beta\sigma}
\end{eqnarray}
on a simple cubic lattice with two orbitals (labeled by $\alpha,\beta$) per site and intra-orbital nearest-neighbor hopping $t$. The Coulomb interaction matrix elements included here are the intra-orbital interaction $U$, the inter-orbital density interaction $V$, and the inter-orbital exchange interaction (Hund's coupling) $J$. The last term represents the transverse part ($S_{i\alpha}^- S_{i\beta}^+ + S_{i\alpha}^+ S_{i\beta}^-$) of the Hund's coupling, and the density interaction term yields the longitudinal part $S_{i\alpha}^z S_{i\beta}^z$, so that altogether the Hund's coupling term has the form $-J {\bf S}_{i\alpha} . {\bf S}_{i\beta}$. The Hamiltonian therefore possesses continuous spin rotation symmetry, and hence the Goldstone mode must exist in the spontaneously broken symmetry state. 

Hund's coupling has been shown to strongly enhance ferromagnetism in an orbitally degenerate system by strongly suppressing the correlation-induced quantum corrections.\cite{hunds} An effective quantum parameter $[U^2 + ({\cal N}-1)J^2]/[U+({\cal N}-1)J]^2$ was obtained for determining the magnitude of quantum corrections for an ${\cal N}$-orbital system, in analogy with $1/S$ for quantum spin systems. The rapid decrease of this quantum factor from 1 to $1/{\cal N}$ as $J/U$ increases from 0 to 1 results in strong suppression of quantum corrections and hence significant stabilization of ferromagnetism by Hund's coupling, particularly for large ${\cal N}$. 

In order to highlight the role of inter-orbital Coulomb interaction $V$ and orbital fluctuations on magnetic couplings and excitations in this paper, we will first set $J=0$. The case of finite Hund's coupling will be treated later in section VII. 

In a band ferromagnet, all information regarding carrier-induced spin interactions $J_{ij} = U^2 \phi_{ij}$
and excitations are contained in the irreducible particle-hole propagator $\phi({\bf q},\omega)$, which then yields the exact transverse spin fluctuation (magnon) propagator:\cite{hunds}
\begin{equation}
\chi^{-+}({\bf q},\omega) = \frac {\phi({\bf q},\omega)}{1- U\phi({\bf q},\omega)} \; .
\end{equation} 
Our approach is to incorporate correlation effects in $\phi({\bf q},\omega)$ in the form of self-energy and vertex corrections using a systematic expansion $\phi = \phi^{(0)} + \phi^{(1)} + \phi^{(2)} + ... $ which preserves the Goldstone mode order by order. Rooted within an inverse-degeneracy expansion scheme, this systematic approach is non-perturbative with respect to the interaction terms and therefore yields a controlled approximation which remains valid in the strong coupling limit. Contributions to the first order quantum correction $\phi^{(1)}$ due to the Hubbard interaction $U$ and Hund's coupling $J$ have been discussed earlier.\cite{hunds}  

\begin{figure}
\vspace*{-10mm}
\hspace*{0mm}
\psfig{figure=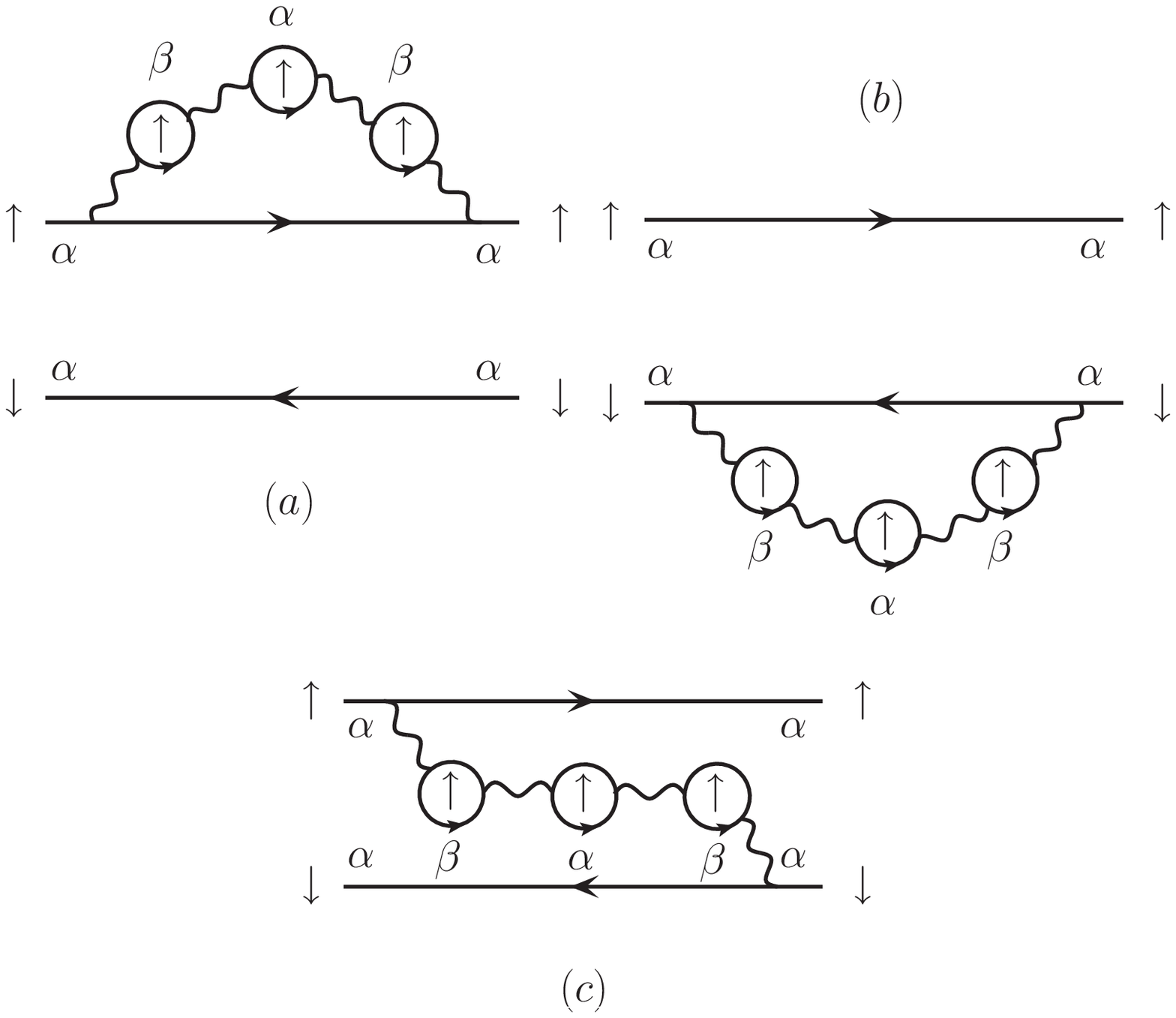,width=80mm}
\psfig{figure=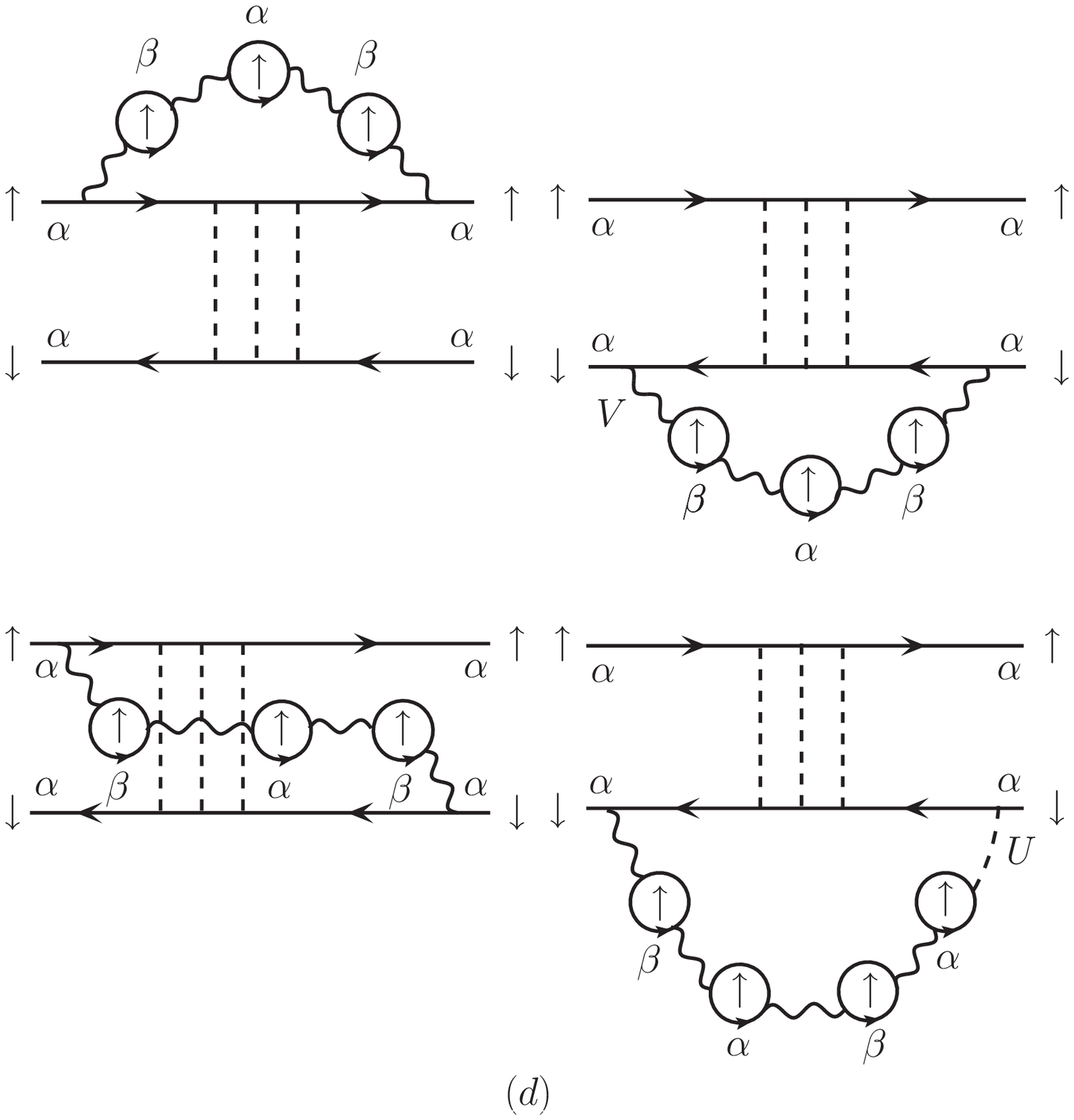,width=80mm}
\caption{First order diagrams for the irreducible particle-hole propagator $\phi$ arising from interaction $V$ involving: (a,b) self energy corrections due to orbital fluctuations, (c) corresponding vertex correction, and (d) vertex corrections involving spin-orbital coupling  (altogether, nine such diagrams). The dashed and wavy lines represent interactions $U$ and $V$, respectively.}
\end{figure}

\begin{figure}
\vspace*{-10mm}
\hspace*{0mm}
\psfig{figure=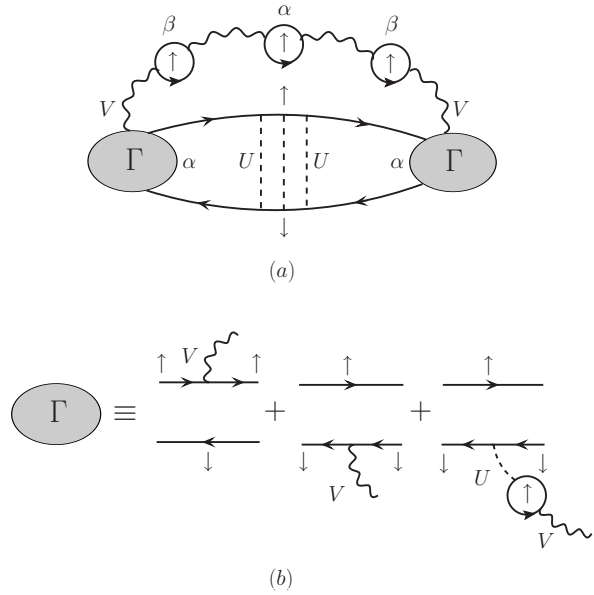,width=80mm}
\caption{The vertex correction diagrams in Fig. 1(d) can be represented (a) in terms of a spin-orbital interaction vertex $\Gamma_{\rm sp-orb}$. The three diagrammatic contributions of $\Gamma_{\rm sp-orb}$ involving three-fermion vertices (b) generate the nine diagrams involving spin-orbital coupling. The missing fourth diagram vanishes because of the assumption of complete polarization.}
\end{figure}

The additional first order diagrams for $\phi$ arising from the inter-orbital interaction $V$ are shown in Fig. 1. The diagrams shown here are for a saturated ferromagnet in which minority ($\downarrow$) spin particle-hole fluctuations are absent. Here Fig. 1(a) and (b) represent quantum corrections due to electronic self energy renormalization by orbital fluctuations, Fig. 1(c) represents the corresponding vertex correction, and Fig. 1(d) represents vertex corrections involving coupling between transverse spin and orbital fluctuations. The vertex correction diagrams as in Fig. 1(d) (nine such diagrams) can be represented in terms of an effective spin-orbital interaction vertex $\Gamma_{\rm sp-orb}$ as shown in Fig. 2(a). The spin-orbital interaction vertex has three contributions involving three-fermion vertices, as shown in Fig. 2(b). The missing fourth diagram vanishes because of the assumption of complete polarization. 

As the Goldstone-mode condition $U\phi=1$ at $q=0$ is already exhausted by the zeroth-order (classical) term $\phi^{(0)}$, the sum of the higher order (quantum) terms $\phi^{(1)}+\phi^{(2)}+ \; ... $ must exactly vanish at $q=0$. For this cancellation to hold for arbitrary $U$, $J$, and $V$, each higher order term $\phi^{(n)}$ in the expansion must individually vanish, implying that the Goldstone mode is preserved order by order. We will demonstrate this exact cancellation explicitly for the new contributions due to $V$ in the first-order quantum correction $\phi^{(1)}$. 

Systematics in our two-orbital model can be formally introduced, in analogy with the inverse-degeneracy ($1/{\cal N}$) expansion for the Hubbard model, by: i) treating the two physical orbitals $\alpha,\beta$ as pseudo spins, ii) introducing ${\cal N}$ pseudo orbitals ($\mu$) for each pseudo spin, and iii) generalizing the inter-orbital density interaction to $(V/{\cal N})\sum_{i\mu\nu} n_{i\alpha\mu} n_{i\beta\nu}$. Now, each interaction line $V$ yields a factor $1/{\cal N}$ and each bubble yields a factor ${\cal N}$ from the summation over pseudo orbitals, resulting in an overall $1/{\cal N}$ factor for the bubble series, and an overall $(1/{\cal N})^n$ factor in the $n^{\rm th}$-order quantum correction $\phi^{(n)}$. 

\section{Spin and Orbital fluctuations}
The diagrammatic expansion above involves spin and orbital fluctuation propagators, the characteristic energy and momentum distribution of which are important in view of the spin-orbital coupling investigated in this work. The ladder series in Fig. 1(d) yields the effective intra-orbital transverse spin interaction:
\begin{equation}
U_{\rm eff}^{\alpha\alpha}({\bf Q},\Omega) = \frac{U}{1-U\chi_0 ({\bf Q},\Omega)} \approx U^2 \frac{\chi_0 ({\bf Q},\Omega)}{1-U\chi_0 ({\bf Q},\Omega)} \equiv U^2 \chi_{\rm sp}({\bf Q},\Omega) 
\approx U^2 \frac{m_{\bf Q}}{\Omega + \omega_{\bf Q}^0 - i \eta} 
\end{equation}
in terms of the RPA-level magnon propagator $\chi_{\rm sp}$, having an advanced pole in the saturated ferromagnetic state ($n_\uparrow = m,\; n_\downarrow = 0$). Here $\chi_0 ({\bf Q},\Omega)$ is the bare antiparallel-spin particle-hole propagator, $m_{\bf Q} \approx m$ and $\omega_{\bf Q}^0$ are the magnon-mode amplitude and energy, and the small weight of gapped Stoner excitations has been neglected for simplicity.

Similarly, the bubble series in Fig. 1 involving odd number of bubbles yields, in terms of the RPA-level orbital fluctuation propagator, the effective intra-orbital density interaction: 
\begin{equation}
V_{\rm eff} ^{\alpha\alpha} ({\bf Q},\Omega)  = -\frac{V^2 \chi_{0\uparrow} ({\bf Q},\Omega)}
{1 - V^2 \chi_{0\uparrow}^2 ({\bf Q},\Omega)} \equiv -V^2 \chi_{\rm orb} ({\bf Q},\Omega) \; \; 
\approx \frac{-V^2}{2} \frac{\chi_{0\uparrow} ({\bf Q},\Omega)}{1 - V \chi_{0\uparrow} ({\bf Q},\Omega)}
\end{equation}
near the orbital ordering instability where $V\chi_{0\uparrow} \sim 1$. Here $\chi_{0\uparrow}({\bf Q},\Omega)$ is the bare spin-$\uparrow$ particle-hole propagator. The orbital fluctuation propagator is symmetric $\chi_{\rm orb} (-{\bf Q},-\Omega) = \chi_{\rm orb} ({\bf Q},\Omega)$, with a spectral representation:
\begin{equation}
\chi_{\rm orb} ({\bf Q},\Omega) = - \int_0^\infty \frac{d\Omega'}{\pi} \frac{{\rm Im} [\chi_{\rm orb} ({\bf Q},\Omega')]}{\Omega - \Omega' + i \eta}
\end{equation}
for its retarded part, with a continuum distribution over the orbital fluctuation energy $\Omega'$. 

Exactly at quarter filling $(m=n=0.5$ per orbital), the orbital fluctuation propagator diverges at ${\bf Q} = (\pi,\pi,\pi)$ in the absence of any NNN hopping $t'$ terms which destroy Fermi surface nesting, indicating instability towards staggered orbital ordering. With increasing hole doping, the spectral function peak shifts below $(\pi,\pi,\pi)$. 

\begin{figure}
\vspace*{-10mm}
\hspace*{0mm}
\psfig{figure=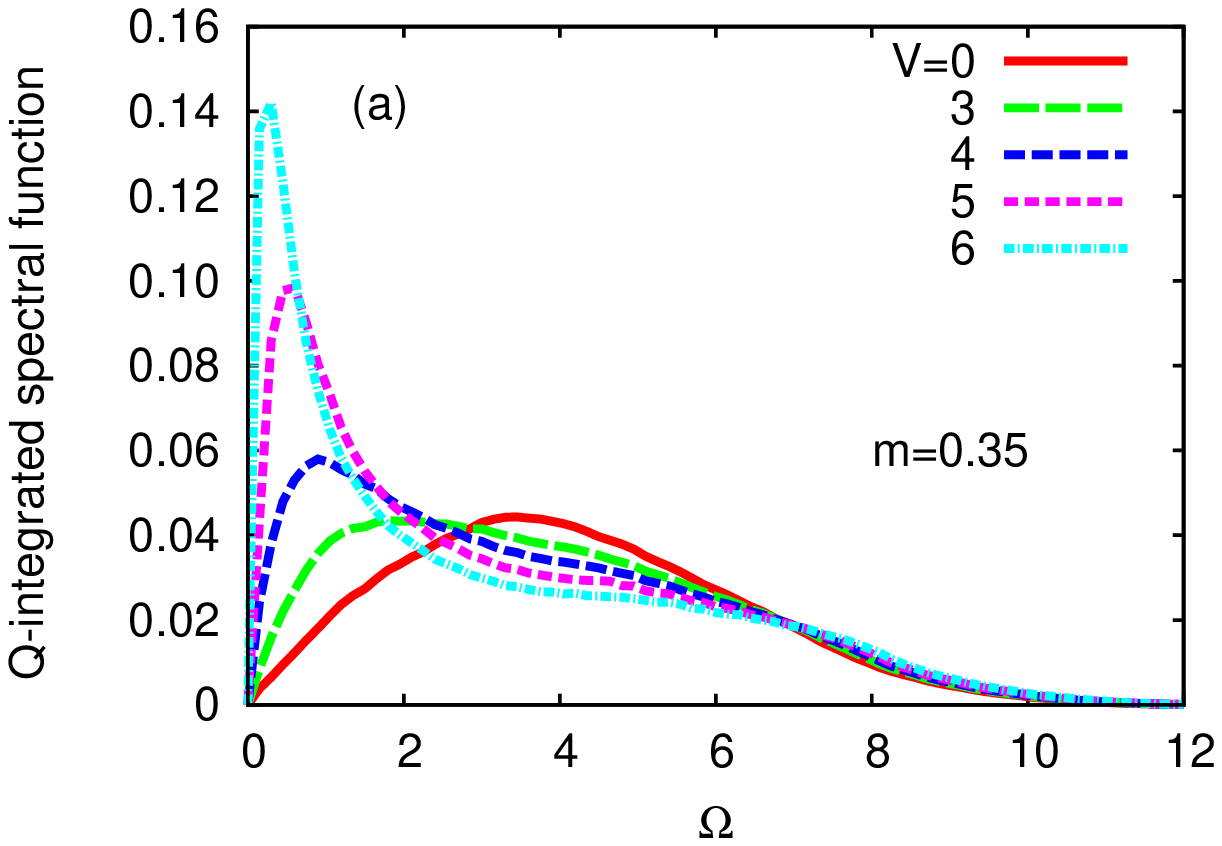,width=80mm}
\psfig{figure=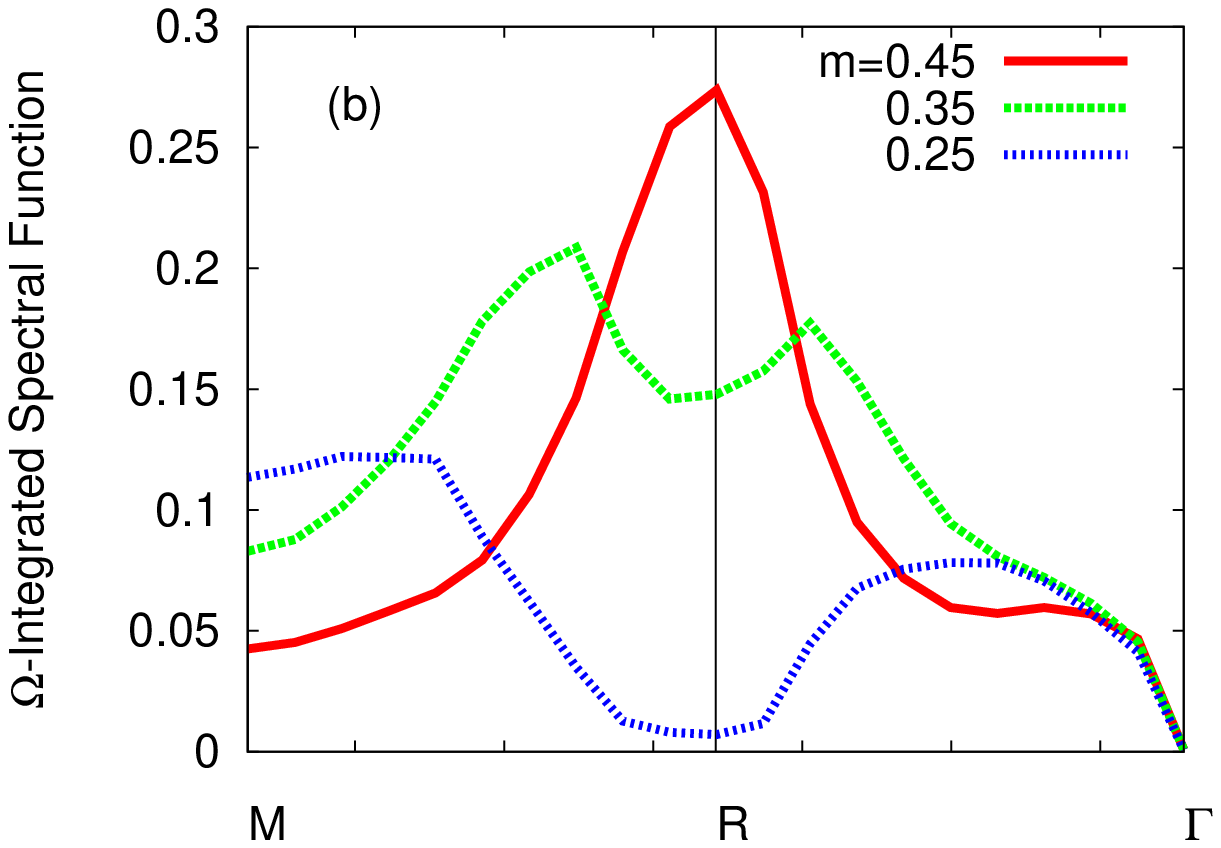,width=80mm}
\caption{Momentum-integrated orbital fluctuation spectral function (a) shows strong suppression of the para-orbiton energy scale near the orbital ordering instability; the low-energy integrated part (b) shows that low-energy orbital fluctuations are concentrated near $(\pi,\pi,\pi)$.} 
\end{figure}

Fig. 3(a) shows the momentum integrated orbital fluctuation (para-orbiton) spectral function 
$\sum_{\bf Q} (1/\pi) {\rm Im} \chi_{\rm orb} ({\bf Q},\Omega)$ with increasing interaction strength $V$. Here, and in the following, we have set the hopping term $t=1=W/12$ as the unit of energy, where $W$ is the bandwidth. In analogy with the well-known para-magnon response with approaching magnetic instability, the para-orbiton energy scale is strongly suppressed from order bandwidth in the weak-coupling regime to relatively very low energies near the orbital ordering instability. 

Fig. 3(b) shows the momentum dependence of the low-energy $(\Omega<2)$ integrated part of the orbital fluctuation spectral function near the R point $(\pi,\pi,\pi)$, which shows that low-energy orbital fluctuations are concentrated near the wavevector $(\pi,\pi,\pi)$ corresponding to staggered orbital fluctuations. With increasing doping away from quarter filling, the peak shifts below $(\pi,\pi,\pi)$, indicating incommensurate fluctuations. 


\section{Orbital fluctuation magnon self energy}
We will first consider diagrams for $\phi$ in Fig. 1 (a,b,c) involving electronic self energy corrections due to orbital fluctuations and the corresponding vertex correction. Absent in the single-orbital case, these diagrams are characteristic of orbital degeneracy, inter-orbital interaction, and orbital fluctuations, and strongly influence magnetic couplings and excitations through electronic band renormalization, particularly in vicinity of the orbital ordering instability. The vertex correction diagrams (d) involving both spin and orbital fluctuations will be discussed in the next section.

Quantum corrections to the irreducible particle-hole propagator $\phi$ in Eq. (2) yield the magnon self energy:  
\begin{equation}
\Sigma ({\bf q},\omega) = mU^2[\phi^{(1)}({\bf q},\omega) + \phi^{(2)}({\bf q},\omega) + ...] \; ,
\end{equation}
in terms of which the magnon propagator 
$\chi^{-+}({\bf q},\omega) =  m/[\omega + \omega_{\bf q} ^{(0)} - \Sigma ({\bf q},\omega)]$. The first-order magnon self energy corresponding to diagrams in Fig. 1(a,b,c) involving only orbital fluctuations is then obtained by summing over the bosonic degrees of freedom of the orbital fluctuations:
\begin{equation}
\Sigma_{\rm orb} ^{(1)} ({\bf q},\omega) = mU^2 [\phi^{(a)} + \phi^{(b)} + \phi^{(c)}] 
= m V^2 \sum_{\bf Q} \int_{-\infty} ^{\infty} \frac{d\Omega}{2\pi i} 
\; \chi_{\rm orb} ({\bf Q-q},\Omega-\omega) \; \Gamma_4 ({\bf Q,q},\Omega,\omega) 
\end{equation}
where $\Gamma_4$ is the four-fermion vertex obtained by integrating out the fermionic degrees of freedom in the diagrams for $\phi$ shown in Fig. 1(a,b,c). As will be further discussed in the following three subsections, this orbital fluctuation magnon self energy physically represents contributions due to (i) coupling between Stoner excitations and orbital fluctuations, and (ii) self-energy corrections involving band-energy renormalization and spectral-weight transfer.  

Using the spectral representation for the orbital fluctuation propagator (retarded part, since $\Gamma_4$ has only advanced poles with respect to $\Omega$), we obtain:
\begin{eqnarray}
\Sigma_{\rm orb} ^{(1)} ({\bf q},\omega) &=& - m V^2 \sum_{\bf Q} \int_{-\infty}^{\infty} \frac{d\Omega}{2\pi i}
\int_0 ^\infty \frac{d\Omega'}{\pi} \frac{{\rm Im} \chi_{\rm orb} ({\bf Q-q},\Omega')}{\Omega-\omega-\Omega' + i\eta} 
\Gamma_4 ({\bf Q,q},\Omega,\omega) \nonumber \\
&=& m V^2 \sum_{\bf Q'} \int_0^\infty \frac{d\Omega'}{\pi} 
{\rm Im} \chi_{\rm orb} ({\bf Q'},\Omega') \; \Gamma_4 ({\bf Q',q},\Omega',\omega) \nonumber \\ 
&=& m V^2 \langle \Gamma_4 ({\bf Q',q},\Omega',\omega) \rangle_{{\bf Q'},\Omega'} \; ,
\end{eqnarray} 
where the average four-fermion vertex $\langle \Gamma_4 \rangle_{{\bf Q'},\Omega'}$ denotes averaging over orbital fluctuation modes ${\bf Q'}\equiv {\bf Q-q}$. Evaluation of the four-fermion vertex $\Gamma_4$, resolved into different contributions corresponding to distinct physical mechanisms, is discussed below. Term by term, the four-fermion vertex $\Gamma_4$ vanishes identically for $q=0$, in accord with the Goldstone mode. 

\subsection{Stoner-orbital coupling}
In the single-orbital case, the magnon self energy due to spin-charge coupling included a Stoner contribution representing coupling of charge excitations with the gapped part of spin excitations.\cite{spch3} In analogy, diagrams Fig. 1(a,b,c) yield contributions which represent a Stoner-orbital coupling:
\begin{equation}
\Gamma_4 ^{\rm Stoner} = U^2 \sum_{\bf k} 
\left ( \frac{1}{\epsilon_{\bf k-Q}^{\downarrow +} - \epsilon_{\bf k}^{\uparrow -} + \Omega - i \eta} \right )
\left ( \frac{1}{\epsilon_{\bf k-q}^{\downarrow +} - \epsilon_{\bf k}^{\uparrow -} + \omega - i \eta}  
- \frac{1}{\epsilon_{\bf k-Q}^{\downarrow +} - \epsilon_{\bf k-Q+q}^{\uparrow} + \omega - i \eta} \right )^2
\end{equation}
in which the first term represents the Stoner excitation mode $({\bf Q},\Omega)$ and the quadratic term is the Stoner-orbital interaction vertex, which involves only magnetic energy denominators. Here $\epsilon_{\bf k}^\sigma = \epsilon_{\bf k} - \sigma \Delta$ are the ferromagnetic state band energies for the two spins in terms of the free-particle energy $\epsilon_{\bf k} = -2t(\cos k_x + \cos k_y + \cos k_z)$ for the sc lattice and the exchange splitting $2\Delta = mU$. The band energy superscripts $+ (-)$ refer to particle (hole) states. There is no restriction on the energy $\epsilon_{\bf k-Q+q}^{\uparrow}$ in Eq. (9) as both particle and hole states contribute. Of the four terms in this quadratic interaction vertex, the two square terms arise from diagrams Fig. 1(a) and (b), while the cross terms are from diagram Fig. 1(c); the characteristic quadratic structure therefore stems from orbital fluctuations renormalizing electrons of both spins, and is clearly absent in the single-orbital case involving spin-down renormalization only. This intrinsic quadratic structure resulting from orbital degeneracy yields a characteristic non-Heisenberg $(1-\cos q)^2$ magnon self energy, resulting in no spin stiffness reduction but strong zone-boundary magnon energy reduction.

\subsection{Electronic band renormalization}
Due to exchange of inter-orbital fluctuations in the diagram Fig. 1(a) involving intermediate spin-$\uparrow$ states, the spin-$\uparrow$ hole (particle) energies are pulled down (pushed up), increasing the particle-hole energy gap, and thereby suppressing the particle-hole propagator $\phi$. Including the corresponding contributions from the vertex correction diagram Fig. 1(c), we obtain the electronic band renormalization contribution:
\begin{equation}
\Gamma_4 ^{\rm band} = - U^2 \sum_{\bf k}
\left ( \frac{1}{\epsilon_{\bf k-Q+q}^{\uparrow +} - \epsilon_{\bf k}^{\uparrow -} + \Omega - \omega} \right ) 
\left ( \frac{1}{\epsilon_{\bf k-q}^{\downarrow +} - \epsilon_{\bf k}^{\uparrow -} + \omega} 
- \frac{1}{\epsilon_{\bf k-Q}^{\downarrow +} - \epsilon_{\bf k-Q+q}^{\uparrow +} + \omega} \right )^2
\end{equation}
involving one charge and two magnetic energy denominators. The finite infinitesimal term $i\eta$ as in Eq. (9) has been dropped for compactness.

\subsection{Spectral weight transfer}
The electronic self energy correction in diagram Fig. 1(a) also results in spectral-weight transfer and redistribution between occupied and unoccupied spin-$\uparrow$ states. However, there is no net change in occupancy and magnetization. The corresponding spectral weight transfer contribution:
\begin{equation}
\Gamma_4 ^{\rm spectral} = - U^2 \sum_{\bf k} 
\left ( \frac{1}{\epsilon_{\bf k-Q+q}^{\uparrow +} - \epsilon_{\bf k}^{\uparrow -} + \Omega - \omega} \right )^2 
\left ( \frac{1}{\epsilon_{\bf k-q}^{\downarrow +} - \epsilon_{\bf k}^{\uparrow -} + \omega} 
- \frac{1}{\epsilon_{\bf k-Q}^{\downarrow +} - \epsilon_{\bf k-Q+q}^{\uparrow +} + \omega} \right )
\end{equation}
involves one magnetic and two charge energy denominators. The first (negative) contribution corresponds to loss of spin-$\uparrow$ hole spectral weight due to transfer to particle states, and the second (positive) contribution corresponds to the reverse process.

\subsubsection*{Cancellation of most singular contributions}
Singular contributions in Eqs. (9-11) for $\Gamma_4$ exactly cancel out. For example, the most singular contribution in Eq. (9) involving two powers of the vanishing energy denominator ($\epsilon_{\bf k-Q}^{\downarrow +} - \epsilon_{\bf k-Q+q}^{\uparrow +} + \omega$) exactly cancels with the corresponding contribution from Eq. (10). Similarly, the next most singular contributions in Eqs. (9-11) also exactly cancel out. 

\begin{figure}
\vspace*{-10mm}
\hspace*{0mm}
\psfig{figure=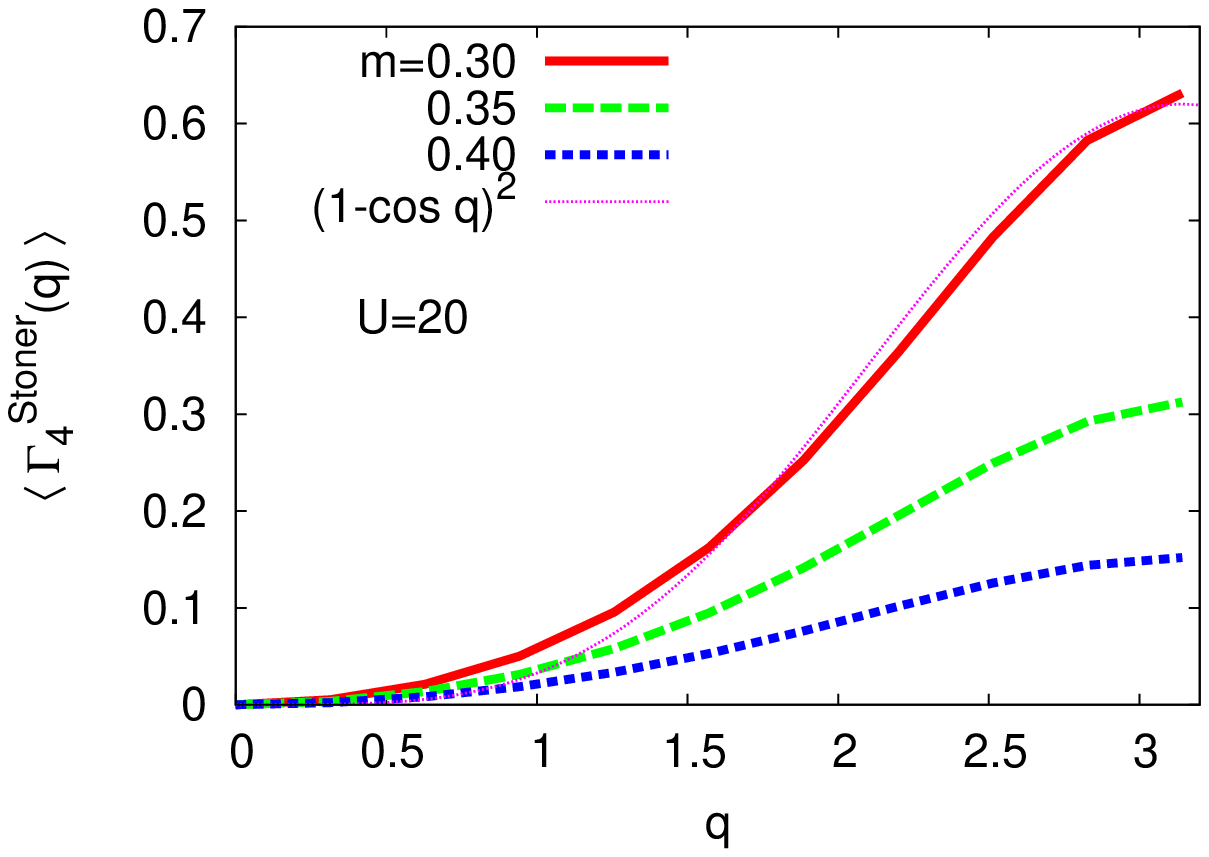,width=52mm}
\psfig{figure=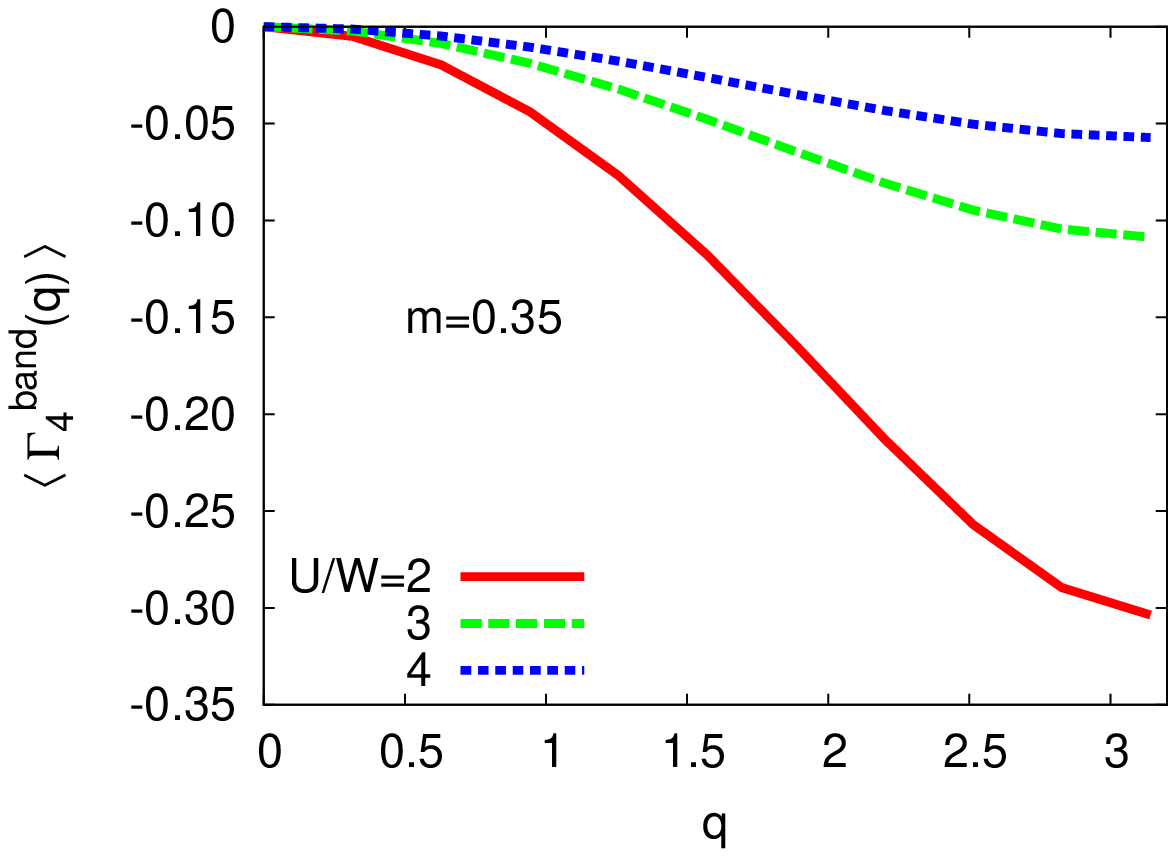,width=52mm}
\psfig{figure=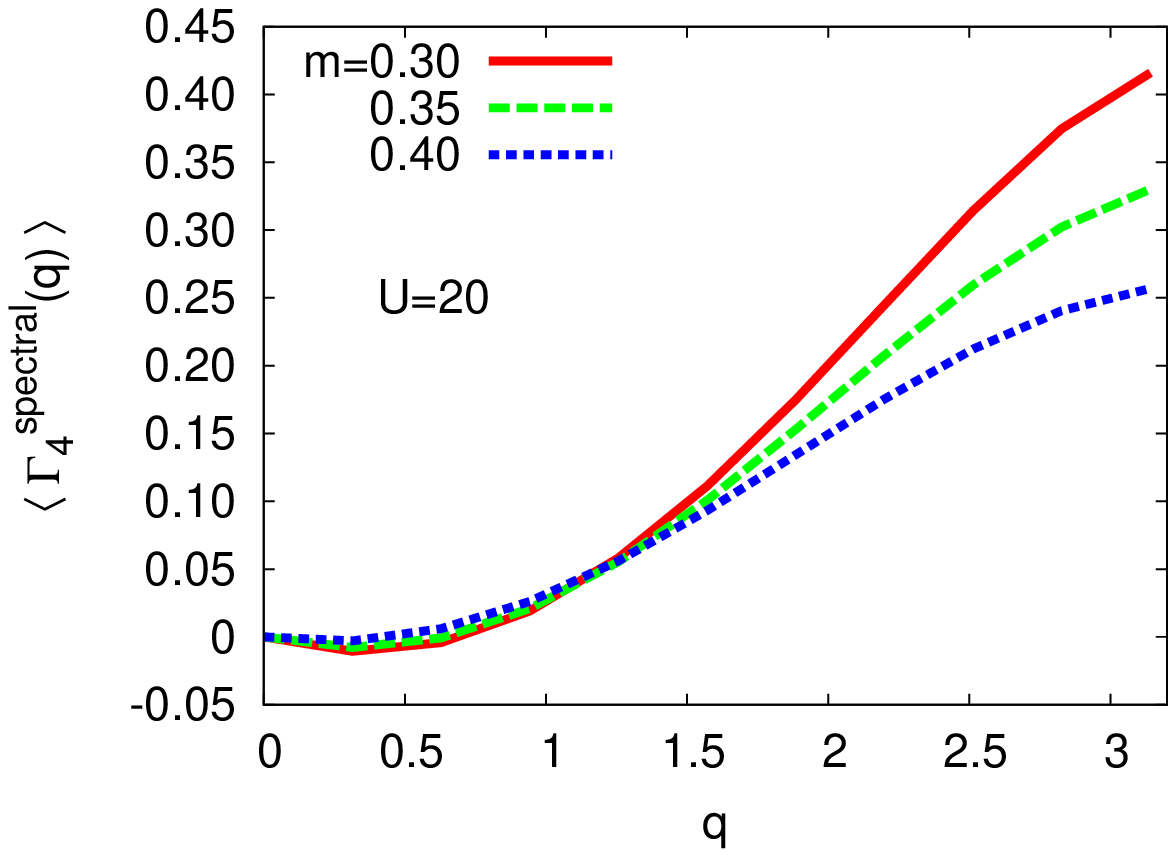,width=52mm}
\vspace{0mm}
\caption{The Stoner, band, and spectral contributions of the averaged four-fermion vertex $\langle \Gamma_4 \rangle$ shows strong non-Heisenberg $(1-\cos q)^2$ behaviour in the $\Gamma$-X direction.}
\end{figure}

\subsubsection*{Average over orbital fluctuation modes}
The average $\langle \Gamma_4 \rangle_{{\bf Q'},\Omega'}$ of the four-fermion vertex over orbital fluctuation modes directly yields the magnon self energy from Eq. (8). Since orbital fluctuations peak below $(\pi,\pi,\pi)$ for finite doping [Fig. 3], the vertex $\langle \Gamma_4 \rangle$ was estimated by averaging over a selected ${\bf Q'}$ region ($-0.8 < \cos Q'_\mu < -0.6$) near $(\pi,\pi,\pi)$, with ${\rm Im} \chi_{\rm orb}({\bf Q'},\Omega')$ assumed flat inside and zero outside. 

The $q$ dependence of the averaged four-fermion vertex $\langle \Gamma_4 (q)\rangle $ in the $\Gamma$-X direction of the Brillouin zone is shown in Fig. 4. The band contribution is negative due to renormalization of band energies by orbital fluctuations, as discussed above. The band contribution rapidly diminishes in the strong coupling limit [Fig. 4(b)], as does the Stoner contribution. The spectral contribution survives in the strong coupling limit. All contributions have strongly non-Heisenberg character, with negligible magnitude at small $q$ but rising sharply near the zone boundary, implying no spin stiffness correction but appreciable zone boundary magnon softening. Orbital fluctuation modes near $(\pi,0,\pi)$ etc. also yield significant non-Heisenberg character to the interaction vertex $\Gamma_4 (q)$, and the three contributions exhibit similar behaviour. 

How does the orbital fluctuation magnon self energy compare with the bare magnon energy? Taking the orbital fluctuation energy $\Omega'$ to be negligible in comparison to the bandwidth near the orbital ordering instability, and the estimated average $\langle \Gamma_4 ({\bf q}) \rangle_{\bf Q'}\approx 0.3$ near the zone boundary from Fig. 4(c), we obtain (for $m=0.35$ and $V=3$):
\begin{equation}
\Sigma_{\rm orb} ({\bf q}) \sim  m (V^2 /2) \langle \Gamma_4 ({\bf q}) \rangle_{\bf Q'} 
= 0.35 \times (9/2) \times 0.3 \approx 0.5 
\end{equation}
which is comparable to the bare zone-boundary magnon energy $\omega_{\bf q}^0 \approx 0.35 (1-\cos q) \approx 0.7$ for realistic strength of the inter-orbital interaction.

\section{Spin-orbital coupling magnon self energy}
In the previous section, we considered the diagrams of Fig. 1 (a,b,c) involving electronic self-energy corrections due to orbital fluctuations. We will now examine the vertex correction diagrams of Fig. 1(d) representing spin-orbital coupling, which are particularly important near the orbital ordering instability where orbital fluctuations are soft. In contrast to the single-orbital case where self energy and vertex correction diagrams were of qualitatively similar order,\cite{singh} introduction of the orbital degree of freedom lifts this constraint and allows qualitatively independent self energy and vertex correction contributions. The corresponding first-order magnon self energy:
\begin{eqnarray}
\Sigma_{\rm sp-orb}^{(1)} ({\bf q},\omega) &=& mU^2 \phi ^{(d)} ({\bf q},\omega) \nonumber \\
&=& -m U^2 \sum_{\bf Q}\int\frac{d\Omega}{2\pi i} [U_{\rm eff} ^{\alpha\alpha} ({\bf Q},\Omega) ] 
[\Gamma_3 ({\bf Q,q},\Omega,\omega)]^2 [V_{\rm eff} ^{\alpha\alpha} ({\bf q-Q},\omega-\Omega) ] \nonumber \\
&=& m \sum_{\bf Q}\int\frac{d\Omega}{2\pi i} [\chi_{\rm sp} ({\bf Q},\Omega) ] 
[\Gamma_{\rm sp-orb} ({\bf Q,q},\Omega,\omega)]^2 [\chi_{\rm orb} ({\bf q-Q},\omega-\Omega) ]
\end{eqnarray}
where $\chi_{\rm sp}$ and $\chi_{\rm orb}$ are the spin and orbital fluctuation propagators (section III), and $\Gamma_{\rm sp-orb} \equiv U^2 V \Gamma_3$ represents the spin-orbital interaction vertex in terms of the three-fermion vertex 
$\Gamma_3$, evaluation of which is discussed in the Appendix.

In analogy with the spin-charge coupling process,\cite{spch3} this correlation-induced spin-orbital coupling is analogous to a second-order Raman scattering process in which the magnon $({\bf q},\omega)$ scatters into an intermediate-state magnon $({\bf Q},\Omega)$ along with an internal orbital excitation $({\bf q-Q},\omega-\Omega)$, leading to significant magnon energy renormalization and magnon damping.

The ${\bf Q},\Omega$ integrals in the above equation represent integrating out the bosonic (both spin and orbital) degrees of freedom. As the magnon propagator $[\chi_{\rm sp} ({\bf Q},\Omega)]$ is purely advanced in nature, only the retarded part of the product $[\Gamma_{\rm sp-orb}]^2 [\chi_{\rm orb}]$ contributes in the $\Omega$ integral. For simplicity, considering the dominant contribution to the spectral representation of this product coming from the orbital fluctuation propagator, from Eqs. (13), (3), (5), and the symmetry property given above Eq. (5), we obtain:
\begin{eqnarray}
\Sigma_{\rm sp-orb}^{(1)} ({\bf q},\omega) 
&=& - m \sum_{\bf Q}\int\frac{d\Omega}{2\pi i} \left ( \frac{m}{\Omega + \omega_{\bf Q}^0 - i \eta} \right )
[\Gamma_{\rm sp-orb} ({\bf Q,q},\Omega,\omega)]^2
\int_0^\infty \frac{d\Omega'}{\pi} \;
\frac{{\rm Im} \chi_{\rm orb}({\bf Q-q},\Omega')}{\Omega - \omega - \Omega' + i \eta}  \nonumber \\
&=& m^2 \sum_{\bf Q} 
\int_0^\infty \frac{d\Omega'}{\pi} \;
\frac {[\Gamma_{\rm sp-orb} ({\bf Q,q},\Omega,\omega)]^2 }
{\omega_{\bf Q}^0 + \omega + \Omega' - i \eta}
{\rm Im}\chi_{\rm orb}({\bf Q-q},\Omega')
\end{eqnarray}


Due to magnon decay into internal spin and orbital excitations ($-\omega=\omega_{\bf q}= \omega_{\bf Q}^0 + \Omega'$), the above magnon self energy yields a finite imaginary part representing finite magnon damping and linewidth, as discussed earlier for spin-charge coupling.\cite{spch3} 

An approximate evaluation of the resulting spin-orbital magnon self energy illustrates the importance of the orbital fluctuation softening near the orbital-ordering instability. 
With $\Omega_{\rm orb}=\Omega'$ and $\Omega_{\rm spin}=\omega_{\bf Q}^0 $ representing characteristic orbital and spin fluctuation energy scales, we obtain (for $\omega=0$): 
\begin{equation}
\Sigma_{\rm sp-orb}^{(1)} ({\bf q}) \approx 
m^2 \frac{\langle [\Gamma_{\rm sp-orb} ({\bf q})]^2 \rangle_{{\bf Q'},\Omega'} }
{\Omega_{\rm spin} + \Omega_{\rm orb} - i \eta }
\end{equation}
where the angular brackets $\langle \; \rangle$ again refer to averaging over the orbital fluctuation modes ${\bf Q' = Q-q}$, as in Eq. (8). Far from the orbital ordering instability, the orbital fluctuation energy $\Omega_{\rm orb}$ is of order bandwidth $W$, which strongly suppresses the magnon self energy. However, near the orbital-ordering instability, spin-orbital coupling becomes important due to the relatively much smaller energy denominator $\Omega_{\rm spin} + \Omega_{\rm orb} \sim t$. 

The spin-orbital interaction vertex $\Gamma_{\rm sp-orb}$ is obtained by integrating out the fermion degrees of freedom in the three-fermion interaction vertices. This interaction vertex explicitly vanishes at momentum $q=0$ in accordance with the Goldstone mode requirement, and yields the dominant $q$ dependence of the magnon self energy. In order to illustrate the characteristic non-Heisenberg character of the interaction vertex, we consider its magnetic part with energy denominators involving the Stoner gap. This term qualitatively differs from the charge part of the vertex with energy denominators involving excitation energies of order bandwidth. Evaluation of the three-fermion vertices contributing to $\Gamma_{\rm sp-orb}$ is discussed in the Appendix. 

For the purely magnetic part, we obtain: 
\begin{equation}
\Gamma_{\rm sp-orb} ^{\rm mag} = U^2 V
\sum_{\bf k}\left ( \frac{1}{\epsilon_{\bf k-Q}^{\downarrow +} - \epsilon_{\bf k}^{\uparrow -} + \Omega - i \eta} \right )
\left [ \frac{1}{\epsilon_{\bf k-q}^{\downarrow +} - \epsilon_{\bf k}^{\uparrow -} + \omega - i \eta} - 
\frac{1}{\epsilon_{\bf k-Q}^{\downarrow +} - \epsilon_{\bf k+q-Q}^{\uparrow} + \omega - i \eta} \right ]
\end{equation}
where there is no restriction on the $\epsilon_{\bf k+q-Q}^{\uparrow}$ energies. 
The leading order $q$ dependence of the above term is approximately $(1-\cos q)$ as shown below. For a fixed orbital fluctuation momentum ${\bf Q'=Q-q}$, with ${\bf Q'}$ close to $(\pi,\pi,\pi)$ corresponding to staggered orbital fluctuations, expansion in powers of $\epsilon/\Delta$ of the square-bracket terms in Eq. (16) yields a contribution $(\epsilon_{\bf k-q} - \epsilon_{\bf k})/2\Delta \sim (1-\cos q)$ to leading order. 
The formally second-order structure of the spin-orbital coupling magnon self energy (Eq. 15) involving $[\Gamma_{\rm sp-orb}]^2$ therefore directly yields an intrinsically non-Heisenberg $(1-\cos q)^2$  contribution to the magnon self energy, which yields no spin stiffness quantum correction, but significant magnon energy reduction and damping near the zone boundary, and therefore accounts for the observed zone-boundary magnon anomalies.

The $q$ dependence of the magnetic part $[\Gamma_{\rm sp-orb} ^{\rm mag}]^2$ of the spin-orbital interaction term is shown in Fig. 5(a). Being small in comparison to the Stoner gap, the boson energies $\omega$ and $\Omega$ in Eq. (16) were set to zero, and the momentum ${\bf Q'}$ was selected in a range near $(\pi,\pi,\pi)$, as in Fig. 4. While $[\Gamma_{\rm sp-orb} ^{\rm mag}]^2$ shows strong anomalous momentum dependence for orbital fluctuation modes near $(\pi,\pi,\pi)$, when both magnetic and charge terms (see Appendix) are included, the net contribution from these modes is small due to a cancellation, as shown in Fig. 5(b). 

\begin{figure}
\vspace*{-10mm}
\hspace*{0mm}
\psfig{figure=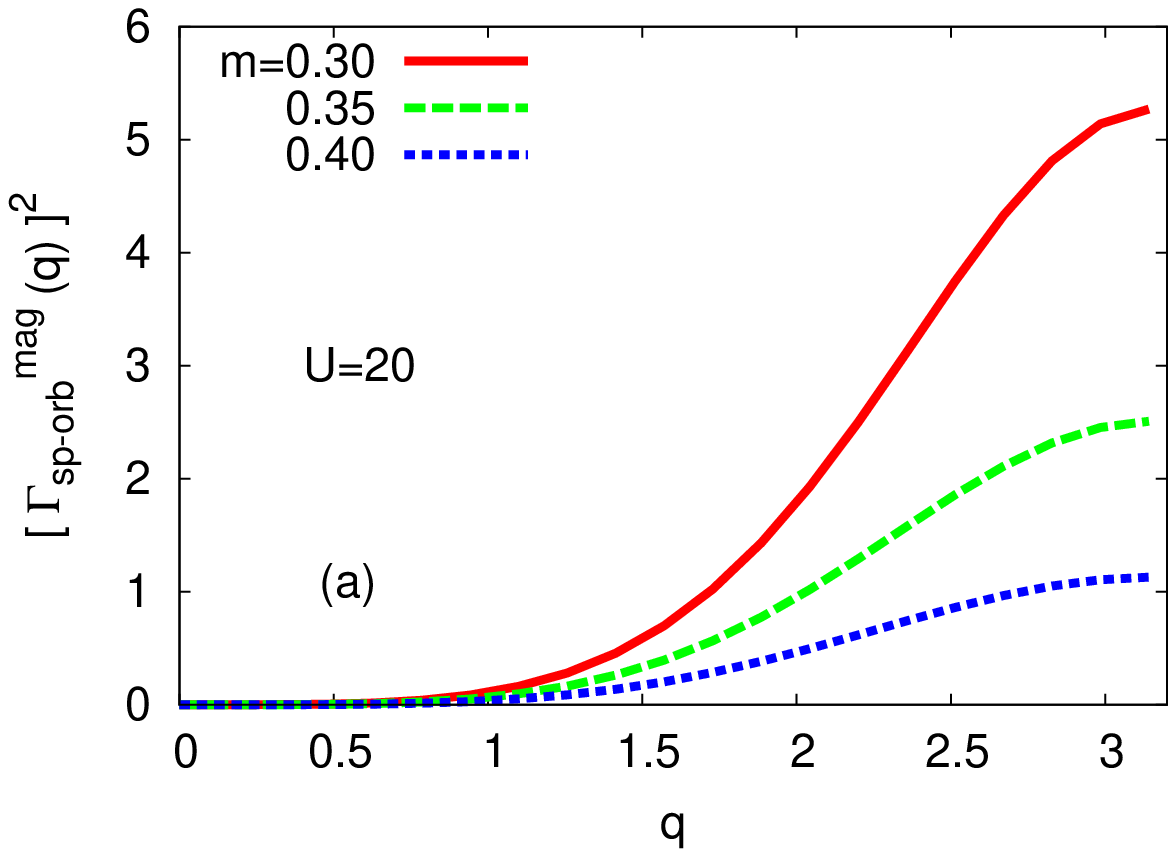,width=80mm}
\psfig{figure=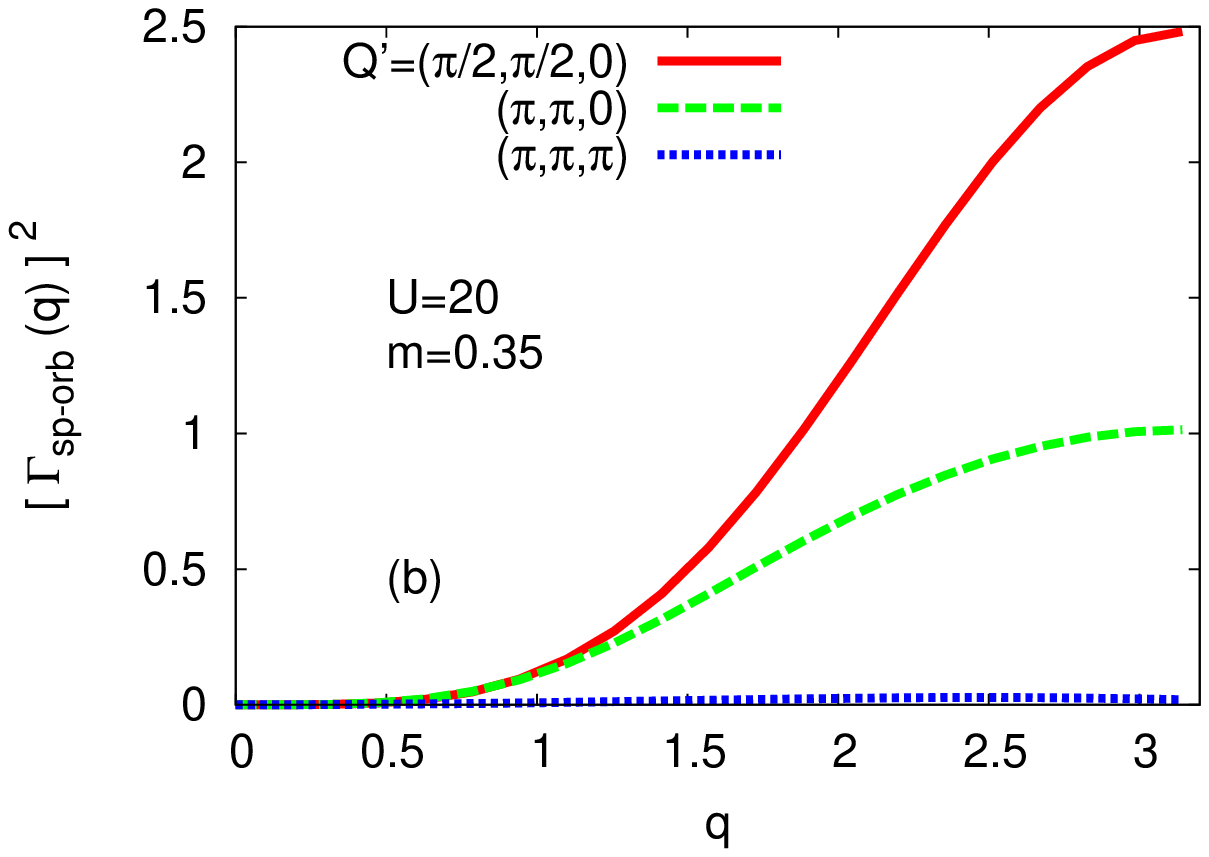,width=80mm}
\vspace{0mm}
\caption{Momentum dependence of (a) the magnetic part $[\Gamma_{\rm sp-orb} ^{\rm mag}]^2$ and (b) the total spin-orbital interaction term $[\Gamma_{\rm sp-orb}]^2$ shows a strong anomalous behaviour in the $\Gamma$-X direction. While staggered orbital fluctuation modes near $(\pi,\pi,\pi)$ yield the dominant contribution to the magnetic part (a), it is modes near $(\pi/2,\pi/2,0)$ which yield the dominant contribution to the total (b), which is also enhanced with hole doping as in (a).}
\end{figure}

\section{Orbital fluctuations near $(\pi/2,\pi/2,0)$}
In contrast, for orbital fluctuation modes with momentum ${\bf Q'}$ near $(\pi/2,\pi/2,0)$, the above cancellation is avoided as the magnetic contribution to $\Gamma_{\rm sp-orb}$ is small. We find that the total spin-orbital interaction term $[\Gamma_{\rm sp-orb}]^2$ exhibits a strong anomalous momentum dependence [Fig. 5(b)], and that it is strongly enhanced with increasing hole doping, as the band filling changes from 0.5 (quarter filling) to 0.25 (one-eighth filling). 

In order to quantitatively examine the effect of this anomalous momentum behaviour of $\Gamma_{\rm sp-orb}^2$ on the magnon dispersion, the magnon self energy was evaluated approximately using Eq. (15). On averaging $[\Gamma_{\rm sp-orb}]^2$ over ${\bf Q'}$, we find that the anomalous momentum behaviour remains qualitatively similar for $|Q_z '| \lesssim 1$ and drops sharply for $Q_z' \gtrsim 1$, whereas in the $Q_x ' - Q_y '$ plane the dominant and qualitatively similar contribution comes from diagonal strips along $|Q_x ' + Q_y '| = \pi$, yielding a factor of $\sim 1/2$ on planar averaging, resulting in an overall phase-space factor of $\sim (1/3)(1/2)$. 

The magnon self energy was therefore estimated using Eq. (15) with $\Gamma_{\rm sp-orb}^2$ as obtained in Fig. 5(b) for the orbital fluctuation mode $(\pi/2,\pi/2,0)$, and including the above phase-space factor (taken as 1/10 in our calculations) to account for momentum averaging over orbital fluctuation modes. Also, the bare magnon energy $\omega_Q ^0$ and the orbital fluctuation energy $\Omega_{\rm orb}$ were taken to be of order $t$. The renormalized magnon energy:
\begin{equation}
\omega_{\bf q} = \omega_{\bf q} ^0 - \Sigma_{\rm sp-orb} ^{(1)} ({\bf q})
\end{equation}
is shown in Fig. 6 for different band fillings $n=(1-x)/2$. It is evident that while the spin stiffness remains essentially unchanged, the magnon self energy at the zone-boundary becomes comparable to the bare magnon energy for realistic values of the inter-orbital interaction $V$, resulting in a zone-boundary softening which is strongly enhanced with hole doping. In the presence of staggered orbital correlations such as near $(\pi/2,\pi/2,0)$, weighted averaging over ${\bf Q'}$ with a peaked orbital spectral function will further enhance $\langle [\Gamma_{\rm sp-orb}]^2 \rangle$ and therefore the anomalous magnon self energy.


\begin{figure}
\vspace*{-10mm}
\hspace*{5mm}
\psfig{figure=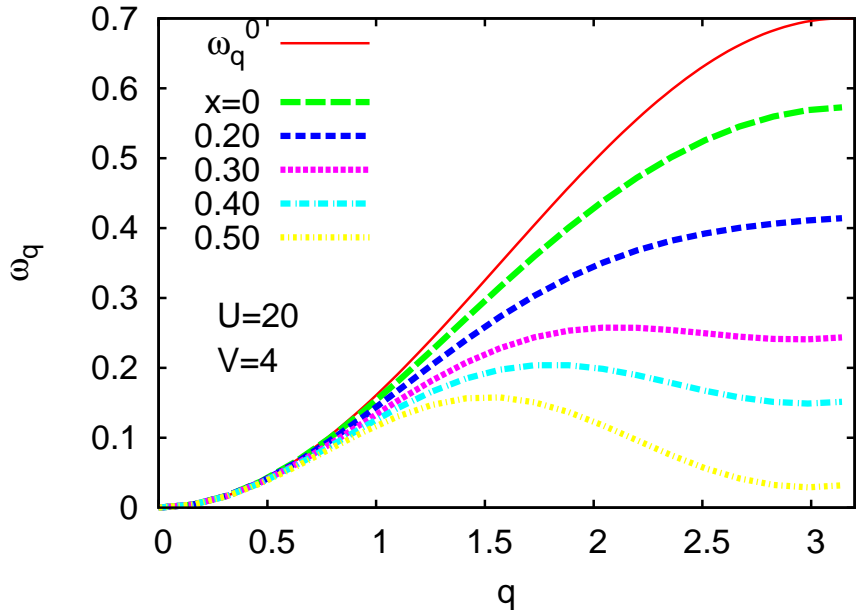,width=120mm}
\vspace{0mm}
\caption{The renormalized magnon energy due to spin-orbital coupling [Eq. (17)] shows strong zone-boundary magnon softening in the $\Gamma$-X direction, which becomes more pronounced with increasing hole doping.}
\end{figure}

The orbital fluctuation modes near $(\pi/2,\pi/2,0)$ correspond to period $4a$ planar staggered correlations, as shown in 
Fig. 7. Such orbital correlations have been observed in the CE-type charge-ordered phase of the half-doped $(x=0.5)$ manganites such as the narrow-band compounds like $\rm Pr_{1-x} Ca_x Mn O_3$ and $\rm La_{1-x} Ca_x Mn O_3$,\cite{kajimoto_2002} and the layered material $\rm La_{1/2}Sr_{3/2}MnO_4$ in which magnetic excitations were found to be dominated by ferromagnetic couplings.\cite{senff_2006} Therefore, as $x$ approaches 0.5, orbital fluctuation modes near $(\pi/2,\pi/2,0)$ may form the dominant contribution to the low-energy part of the orbital fluctuation spectral function. 

\begin{figure}
\vspace*{-10mm}
\hspace*{5mm}
\psfig{figure=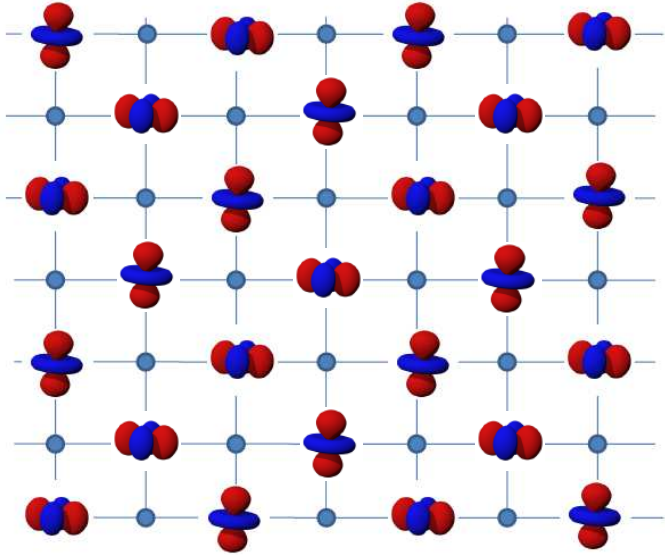,width=120mm}
\vspace{-20mm}
\caption{Period $4a$ planar staggered orbital correlations corresponding to orbital fluctuation modes with momentum near $(\pi/2,\pi/2,0)$.}
\end{figure}

The different layers in $\rm La_{1/2}Sr_{3/2}MnO_4$ are magnetically decoupled due to negligible interlayer couplings. Extension of the present investigation of spin-orbital coupling to the two-dimensional case is therefore of interest with respect to renormalization of magnetic couplings. In the case of spin-charge coupling, the renormalized magnon dispersion for a square lattice does show strong zone-boundary softening near $(\pi,0)$ and $(0,\pi)$, while the magnon energy near $(\pi,\pi)$ remains undiminished, indicating softening of the nearest-neighbour ferromagnetic bonds but strong ferromagnetic correlations along the diagonal directions.\cite{spch3}   


Orbital fluctuation modes near ${\bf Q'}=(\pm\pi,0,0)$ and $(0,\pm\pi,0)$ were also found to yield significant contribution to the spin-orbital interaction vertex. Involving similar density $\langle n_{i\alpha} + n_{i\beta} \rangle$ on all sites, such configurations should, however, be relatively suppressed by an intersite density interaction $V'n_i n_j$. In contrast, allowing for reduced density on alternating ``empty'' sites with vanishing orbital ``magnetization'', configurations corresponding to modes near $(\pi/2,\pi/2,0)$ with period $4a$ orbital correlations minimize the intersite interaction energy $V' n_i n_j$, and would therefore be relatively more important.

\section{Finite Hund's coupling}
So far we had set the Hund's coupling $J=0$ in order to highlight the role of inter-orbital interaction and fluctuation. For finite $J$, it is convenient to proceed in two steps. The part $V=J$ of the inter-orbital interaction $V$ together with Hund's coupling effectively amounts to a purely magnetic interaction $-J{\bf S}_{i\alpha}.{\bf S}_{i\beta}$, and has been investigated earlier.\cite{hunds} This is because for $V=J$, the inter-orbital interaction $(J-J\delta_{\sigma\sigma'})n_{i\alpha\sigma} n_{i\beta\sigma'}$ acts only between opposite-spin electrons, and so the resulting diagrammatics is similar to the Hubbard interaction case. The remaining part $(V-J)$ is purely non-magnetic, and yields diagrammatic contributions essentially as in section II. Thus, corresponding to Fig. 1 (a,b,c) diagrams involving the bubble series for the orbital fluctuation propagator, the magnon self energy is obtained by simply replacing $V$ by $(V-J)$ in Eq. (7). 

For the spin-orbital coupling magnon energy, however, the interaction ladders in diagrams Fig. 1(d) and Fig. 2 now involve Hund's coupling  $J$ as well, and since the transverse part of $J$ flips the orbital index in the ladder series, there are now two contributions to the irreducible particle-hole propagator:
\begin{equation}
\phi_{\alpha\mu}({\bf q},\omega) = - \sum_{\bf Q} \int \frac{d\Omega}{2\pi i} 
U_{\rm eff} ^{\alpha\mu} ({\bf Q},\Omega) 
[\Gamma_3 ({\bf Q,q},\Omega,\omega)]^2 
V_{\rm eff} ^{\alpha\mu} ({\bf q-Q},\omega-\Omega) 
\end{equation}
involving intra $(\mu=\alpha)$ and inter $(\mu=\beta)$ orbital spin and orbital fluctuations. Involving ladders of $U$ and $J$, the effective intra and inter-orbital transverse spin interactions:
\begin{equation}
U_{\rm eff} ^{\alpha\mu} = \frac{1}{2} \left [\frac{U^+}{1-U^+ \chi_0} \pm \frac{U^-}{1-U^- \chi_0} \right ] 
\approx \frac{(U+J)^2}{2} \left [\frac{\chi_0}{1-U^+ \chi_0} \pm \frac{\chi_0}{1-U^- \chi_0} \right ]
\end{equation}
can be expressed as in-phase $(\mu=\alpha)$ and out-of-phase $(\mu=\beta)$ combinations of the acoustic and optical branches,\cite{hunds} with $U^{\pm} \equiv U \pm J$. Similarly, the effective intra and inter-orbital density interactions:
\begin{eqnarray}
V_{\rm eff} ^{\alpha\alpha} &=& -\frac{(V-J)^2\chi_{0\uparrow}}{1 - (V-J)^2\chi_{0\uparrow} ^2} \nonumber \\
V_{\rm eff} ^{\alpha\beta}  &=& \frac{(V-J)}{1 - (V-J)^2\chi_{0\uparrow} ^2}
\end{eqnarray}
involve odd and even number of bubbles.

Now, in the investigation of role of Hund's coupling on quantum corrections,\cite{hunds} it was shown that the inter-orbital component $U_{\rm eff} ^{\alpha\beta} ({\bf Q},\Omega)$ yields a small $\Omega$-integrated contribution due to partial cancellation from the out-of-phase combination of the acoustic and optical modes, essentially reflecting an inter-orbital incoherence. Therefore, the inter-orbital component $\phi_{\alpha\beta}$ should be relatively small, and the intra-orbital component $\phi_{\alpha\alpha}$ is approximately given by Eq. (15), with $V$ replaced by $(V-J)$ and $U$ replaced by $(U+J)$. 

Due to the purely opposite-spin density interaction $(J-J\delta_{\sigma\sigma'})n_{i\alpha\sigma} n_{i\beta\sigma'}$, the interaction line $U$ connected to the bubble in the third diagram in Fig. 2(b) for the three-fermion vertex $\Gamma_3$ is also replaced by $(U+J)$. This is consistent with the enhanced exchange splitting to $(U+J)m$ in the $\chi^0$ energy denominator, which changes the magnon pole condition to $(U+J)\chi^0 = 1$, and ensures that the three-fermion vertex $\Gamma_3$ exactly vanishes at $q=0$ in accordance with the Goldstone-mode condition (see Appendix). 

\section{Extension to the ferromagnetic Kondo lattice model}
Magnetic couplings and excitations in ferromagnetic manganites have been theoretically investigated using the ferromagnetic Kondo lattice model (FKLM) and its strong coupling limit, the double exchange ferromagnet. In this model, the $S=3/2$ core spins due to localized $\rm t_{2g}$ electrons of the magnetic Mn$^{++}$ ions are exchange coupled to the mobile $\rm e_g$ band electrons, represented by an interaction term $-{\cal J} \sum_i {\bf S}_i . {\mbox{\boldmath $\sigma$}}_i$, with $\cal J$ of order bandwidth $W$. The fermionic representation approach for evaluating magnon self energy corrections in the FKLM, which allows conventional diagrammatic tools to be employed for evaluating quantum corrections beyond the leading order,\cite{qfklm} can be readily extended to include effects of orbital fluctuations. 

With an inter-orbital interaction $V$ included between the two degenerate $\rm e_g$ orbitals, the FKLM magnon self energy due to spin-orbital coupling can be directly obtained from our magnon renormalization analysis of Fig. 6. The required correspondence is: $U\rightarrow {\cal J}$ and $m\rightarrow 2S$, so that the exchange splitting $Um \rightarrow 2{\cal J}S$. The FKLM magnon self energy is thus obtained from our calculated Hubbard model result using a multiplicative factor $f={\cal J}^4(2S)^2/U^4 m^2$. With $U/t=20$, $m=0.35$, ${\cal J}/t=4$, and $2S=3$, we obtain $f\approx 1/6$. As the FKLM bare magnon energy $\sim (t/18)(1-\cos q)$ is smaller than the Hubbard model bare magnon energy $\sim (t/3)(1-\cos q)$ by roughly the same factor, the FKLM renormalized magnon energy $\omega_{\bf q}=\omega_{\bf q}^0 - \Sigma_{\rm sp-orb} ({\bf q})$ will also be as in Fig. 6, only scaled down by a factor (1/6). Taking the hopping energy scale $t\sim 200$meV corresponding to a realistic bandwidth $\sim 2$eV, the magnon energy scale in Fig. 6 is $\sim 30$meV for ferromagnetic manganites, in agreement with the measured magnon energies.\cite{zhang_2007}  

Since the FKLM magnon self energy goes as the fourth power of ${\cal J}/t$ explicitly, the anomalous zone-boundary softening effect should be especially pronounced in narrow-band systems. Indeed, zone-boundary magnon softening is clearly seen to occur in the relatively low-$T_{\rm C}$ or narrow-band materials, and broad-band materials such as $\rm La_{0.7} Pb_{0.3} Mn O_3$ show nearly Heisenberg behavior, in agreement with this prediction. Systematic studies of doping dependence of the zone-boundary magnon softening indicates that the higher the doping level, the larger the zone-boundary softening.\cite{zhang_2007} Furthermore, doping dependence of spin dynamics indicates that the measure spin stiffness $D \sim 160 \pm 15$ meV \AA$^2$ remains essentially unchanged, while the zone-boundary magnon softening (denoted by the ratio $J_4/J_1$) is found to be enhanced linearly with increasing doping.

\section{Conclusions}
The correlated motion of electrons in the presence of strong orbital fluctuations and correlations was investigated in an orbitally degenerate ferromagnet with two orbitals per site with respect to magnetic couplings and excitations. A systematic Goldstone-mode-preserving approach was employed to incorporate correlation effects in the form of self energy and vertex corrections, so that both long-wavelength and zone-boundary magnon modes could be studied on an equal footing. Our investigation focussed on the anomalous momentum dependence of the three- and four-fermion interaction vertices which determine the magnon self energy and the role of different orbital fluctuation modes, particularly near the orbital ordering instability where orbital fluctuations are relatively soft.

Orbital fluctuations were generically shown to impart an intrinsically non-Heisenberg $(1- \cos q)^2$ character to the magnon self energy in the $\Gamma$-X direction of interest. This generic behaviour was ascribed to a quadratic structure of the spin-orbital interaction vertex resulting from a new class of diagrammatic contributions associated with the orbital degree of freedom. These diagrams are absent in the single-orbital case, the essential difference being that orbital fluctuations couple to electrons of both spin. 

The absence of any $q^2$ contribution in this non-Heisenberg magnon self energy for small $q$ implies that the spin stiffness is not renormalized by orbital fluctuations generated by the inter-orbital density interaction $V$. In a multi-orbital ferromagnet, with intra-orbital Coulomb interaction $U$, inter-orbital interaction $V$, Hund's coupling $J$, and orbital degeneracy ${\cal N}$, the spin stiffness therefore continues to be essentially determined by the intra-atomic factors $U$, $J$, ${\cal N}$, through the effective quantum parameter $[U^2 + ({\cal N}-1)J^2]/[U + ({\cal N}-1)J]^2$ as obtained earlier,\cite{hunds} and the interaction $V$ does not play an important role. 

However, the strong enhancement of magnon self energy near the zone boundary resulted in a strong anomalous magnon softening in the $\Gamma$-X direction, which increases significantly with hole doping away from quarter filling. Our investigation thus clarifies the completely different roles of interactions $J$ and $V$, representing magnetic and charge parts of the inter-orbital Coulomb interaction. While Hund's coupling enhances ferromagnetism by strongly suppressing the effective quantum parameter, orbital fluctuations and correlations due to $V$ destabilize ferromagnetism by strongly suppressing zone-boundary magnon energies near the orbital ordering instability.   

With regard to relative importance of different orbital fluctuation modes, staggered fluctuations with ${\bf Q}$ near $(\pi,\pi,\pi)$ and $(\pi,0,\pi)$ were found to be most important for the orbital fluctuation magnon self energy. The spin-orbital coupling magnon self energy was found to be strongly sensitive to orbital fluctuation modes due to an interference between magnetic and charge terms in the interaction vertex. Thus, while the magnetic part showed strong anomalous momentum dependence for orbital modes near $(\pi,\pi,\pi)$, the net contribution to the total vertex was found to be small due to cancellation with the charge term. Rather, fluctuation modes near $(\pi/2,\pi/2,0)$ were found to be important for the total vertex including the charge part. The strong zone-boundary magnon softening near $(0,0,\pi)$, arising from staggered orbital fluctuations with ${\bf Q}$ near $(\pi/2,\pi/2,0)$, suggests an instability towards a composite structure of spin-orbital correlations involving period $4a$ orbital ordering in ferromagnetic planes and intra-orbital AF spin correlations in the perpendicular direction. 

These results provide a plausible explanation of the observed anomalies in neutron scattering studies of spin-wave excitations in ferromagnetic manganites, where spin stiffness is seen to remain essentially unchanged whereas the zone-boundary magnon softening is enhanced with increasing hole doping and the approach towards CE-type charge-orbital ordered states near $x=0.5$. Our results of strong anomalous magnon self energy contribution from different orbital fluctuation modes such as $(\pi,\pi,\pi)$ and $(\pi/2,\pi/2,0)$ show the zone-boundary softening to be a more generic feature of spin-orbital coupling. Only ferromagnetic orbital correlations extremely close to the orbital ordering instability were found to yield significant magnon softening in earlier studies.\cite{khaliullin_2000}  

The observed zone-boundary magnon softening has been usually modeled by including a fourth neighbour interaction term $J_4$.\cite{zhang_2007} As $J_4$ yields no contribution to the zone-boundary magnon energy, but contributes significantly to spin stiffness, it must be accompanied by a corresponding reduction $\Delta J_1 = 4J_4$ in the NN coupling so that the spin stiffness remains unchanged, as observed experimentally; the net magnon energy reduction then has the non-Heisenberg form $2J_4 (1-\cos q)^2$. Our anomalous magnon self energy result of this form thus provides fundamental insight into the role of orbital fluctuations on magnetic couplings and excitations. 


Instead of the inter-orbital interaction $Vn_{i\alpha} n_{i\beta}$ considered here, inter-site interactions $V' n_i n_j$ would generate similar diagrammatic contributions to the magnon self energy, which become important near the charge-ordering instability where charge excitations become relatively soft, resulting in similar magnon self energy and anomalous zone-boundary magnon softening. 


\newpage
\section*{Appendix}
The spin-orbital interaction vertex is obtained as: 
\begin{equation}
\Gamma_{\rm sp-orb} = U^2 V \left [ \Gamma_3^{(a)} + \Gamma_3^{(b)} + \Gamma_3^{(c)} \right ]
\end{equation}
in terms of the three types of fermion vertices shown in Fig. 1(e), which are evaluated by integrating out the fermion degrees of freedom as discussed below.

For the first fermion vertex with interaction line $V$ attaching to spin-$\downarrow$ fermion, we obtain:
\begin{equation}
\Gamma_3^{(a)} = -\sum_{\bf k} \left ( \frac{1}{\epsilon_{\bf k-q}^{\downarrow +} - \epsilon_{\bf k}^{\uparrow -} + \omega - i \eta} 
\right ) \left ( \frac{1}{\epsilon_{\bf k-Q}^{\downarrow +} - \epsilon_{\bf k}^{\uparrow -} + \Omega - i \eta} \right ) \; .
\end{equation}
The second vertex $\Gamma_3^{(b)} = \Gamma_3^{(b1)} + \Gamma_3^{(b2)}$ consists of a similar all magnetic term: 
\begin{equation}
\Gamma_3^{(b1)} = \sum_{\bf k} \left ( \frac{1}{\epsilon_{\bf k-Q}^{\downarrow +} - \epsilon_{\bf k-Q+q}^{\uparrow } + \omega - i \eta} \right ) \left ( \frac{1}{\epsilon_{\bf k-Q}^{\downarrow +} - \epsilon_{\bf k}^{\uparrow -} + \Omega - i \eta} \right )
\end{equation}
and a magnetic-charge term:
\begin{eqnarray}
\Gamma_3^{(b2)} &=& - \sum_{\bf k} \left [\left ( \frac{1}{\epsilon_{\bf k-q}^{\downarrow +} - \epsilon_{\bf k}^{\uparrow -} + \omega - i \eta} \right )\left ( \frac{1}{\epsilon_{\bf k-q+Q}^{\uparrow +} - \epsilon_{\bf k}^{\uparrow -} + \omega - \Omega - i \eta} \right ) \right . \nonumber \\ 
& + & \left . \left ( \frac{1}{\epsilon_{\bf k-Q}^{\downarrow +} - \epsilon_{\bf k-Q+q}^{\uparrow +} + \omega - i \eta} \right ) \left ( \frac{1}{\epsilon_{\bf k-Q+q}^{\uparrow +} - \epsilon_{\bf k}^{\uparrow -} - \omega + \Omega - i \eta} \right ) \right ] \; .
\end{eqnarray}
In Eq. (23), there is no restriction on the fermion energy $\epsilon_{\bf k-Q+q}^{\uparrow }$ as contributions with both particle (+) and hole (-) energies are included. Finally, the third vertex $\Gamma_3^{(c)}$ involves a $U$ interaction line and a spin-$\uparrow$ bubble attached to the spin-$\downarrow$ fermion lines, and is given by: 
\begin{equation}
\Gamma_3^{(c)} = - \Gamma_3^{(a)} \; U \; \chi_{0\uparrow}({\bf q-Q},\omega-\Omega)
\end{equation} 

For $q=0$, the spin-orbital interaction vertex $\Gamma_{\rm sp-orb}$ vanishes identically, ensuring that the Goldstone mode is explicitly preserved. Both $-\Gamma_3^{(a)}$ and $\Gamma_3^{(b1)}$ reduce to $\chi_0({\bf Q},\Omega)/(2\Delta +\omega)$ for $q=0$, whereas both $-\Gamma_3^{(b2)}$ and $\Gamma_3^{(c)}$ reduce to $\chi_{0\uparrow}({\bf Q},\omega-\Omega)$ on setting $U\chi_0({\bf Q},\Omega)=1$ at the magnon pole, so that from Eq. (21) $\Gamma_{\rm sp-orb}=0$ for $q=0$.

Singular contributions in the three-fermion vertex $\Gamma_3$ exactly cancel out, as for the four-fermion vertex $\Gamma_4$ discussed below Eq. (11). Thus, the singular contribution of Eq. (23) due to vanishing energy denominator $\epsilon_{\bf k-Q}^{\downarrow +} - \epsilon_{\bf k-Q+q}^{\uparrow +} + \omega$ exactly cancels the corresponding contribution from the second term of Eq. (24).


\end{document}